\documentclass[11pt, oneside]{article}   	

\usepackage{chemformula} 
\usepackage[T1]{fontenc} 

\usepackage{geometry}                		
\usepackage[parfill]{parskip}    		
\usepackage{graphicx}				
\usepackage{amssymb}
\usepackage{bm}
\usepackage{siunitx}
\usepackage{booktabs}

\def\D{{{\mathcal{D}}}}
\def\Dc{\D}
\def\Fc{{{\mathcal{F}}}}
\def\uv{{\mathbf{u}}}
\def\Rv{{\mathbf{R}}}
\def\rv{{\mathbf{r}}}

\def\csch{{\mbox{csch}}}

\def\vv{{\mathbf{v}}}
\def\qv{{\mathbf{q}}}
\def\kv{{\mathbf{k}}}
\def\O{\Omega}

\title{Compression-induced continuous phase transition in the buckling of a semiflexible filament for two and three dimensions}
\author{Ananya Mondal$^{1,2}$ and Greg Morrison$^{1,2,*}$\\
{$^1$ Department of Physics, University of Houston, Houston TX 77204.}\\
{$^2$ The Center for Theoretical Biological Physics, Houston TX 77030.}\\
{$^*$ gcmorrison@uh.edu}
}

\begin{document}
\maketitle

\begin{abstract}
The ability of biomolecules to exert forces on their surroundings or resist compression from the environment is essential in a variety of biologically relevant contexts.  As has been understood for centuries, slender rods can only be compressed so far until they buckle, adopting an intrinsically bent state that may be unable to bear a compressive load.  In the low-temperature limit and for a constant compressive force, Euler buckling theory predicts a sudden transition from a compressed to a bent state in these slender rods.  In this paper, we use a mean field theory to show that if a semiflexible chain is compressed at finite temperature with fixed end-to-end distance (permitting fluctuations in the compressive forces), it exhibits a continuous phase transition to a buckled state at a critical level of compression, and we determine a quantitatively accurate prediction of the transverse position distribution function of the midpoint of the chain that indicates the transition. We find the mean compressive forces are non-monotonic as the extension of the filament varies, consistent with the observation that strongly buckled filaments are less able to bear an external load.  We also find that for the fixed extension (isometric) ensemble that the buckling transition does not coincide with the local minimum of the mean force (in contrast to Euler buckling).  We also show the theory is highly sensitive to fluctuations in length, and that the buckling transition can still be accurately recovered by accounting for those fluctuations.  These predictions may be useful in understanding the behavior of filamentous biomolecules compressed by fluctuating forces,  relevant in a variety of biological contexts.  
\end{abstract}

\section{Introduction}

Buckling is defined as the process by which a slender column bends laterally under an axial compressive load. At the cellular level, buckling instabilities have been observed \cite{murrell2015forcing, costa2002buckling, bathe2008cytoskeletal, enrique2015actin, bashirzadeh2020confinement, bashirzadeh2021actin} to occur in biological systems as well with cytoskeletal filaments like F-actin which buckle during cell deformation. The ability of a single or bundled cytoskeletal filaments to exert forces, support an external load, or deform membranes is essential for many biological processes \cite{borovac2018regulation, runge2020dendritic, jacquemet2015filopodia, calzado2016effect, mogilner2005physics,mellor2010role, bornschlogl2013filopodia, katoh1999arrangement}, and it is important to relate the mechanical properties of the filaments to their ability to perform these functions.  Deformation forces resulting from cell shape changes can also lead to internal re-organization of cross-linked actin bundles \cite{lieleg2008transient, lieleg2010structure, broedersz2014modeling, mulla2019origin, strehle2011transiently}, and can induce buckling in some cases. The ends of actin filaments inside a cell can be mechanically coupled to other biomolecules forming the cytoskeletal network \cite{brangwynne2006microtubules, chaudhuri2007reversible, freedman2017versatile, berro2007attachment, martiel2020force, khanduja2014processive, xu2022modeling} and these attachment conditions can generate contractile forces inducing buckling. Other more complex cases of buckling inside a cell arise from actomyosin contractility \cite{lenz2012contractile, ronceray2016fiber, wollrab2019polarity, morris2022synaptopodin, komianos2018stochastic} involving ATP-dependent compression by myosin-motors, or even axial compression on the actin bundles by cell membrane tension \cite{walani2015endocytic, simon2019actin, ni2021membrane}. Actomyosin contractility can also lead to many interesting phenomena such as helical buckling \cite{tanaka1992super, leijnse2015helical, leijnse2022filopodia} or development of sharp kinks \cite{schramm2017actin, vogel2013myosin, deguchi2012simultaneous, okamoto2020helical, jung2016f,li2017buckling, murrell2012f, koenderink2018architecture, murrell2012f} that eventually cause sharply bent filaments to sever.

In addition to experimental studies of buckling $in\  vivo$, numerous $in\  vitro$ studies have probed the responses of semiflexible biomolecules to compressive forces using single-molecule force experiments for single filaments \cite{ferrer2008measuring, rief2000myosin, stuij2019stochastic, footer2007direct, berro2007attachment, kovar2004insertional, forth2008abrupt}, cross-linked bundles \cite{ruckerl2017adaptive, claessens2006actin}, and networks \cite{lehmann2020optical, broedersz2014modeling, murrell2012f}. Cross-linked actin networks exhibit strain-stiffening \cite{kang2009nonlinear, gardel2004scaling, gardel2008mechanical, haase2015investigating} a characteristic property of semiflexible polymers which show greater resistance to elongation the more they are stretched. At high strains, part of the network also experiences compression and actin filaments tend to buckle \cite{amuasi2015nonlinear} leading to severing. As a model to explore the role of deformation forces on buckling in the actomyosin cell cortex, artificially made giant unilamellar vesicles (GUVs) are used \cite{liu2008membrane, murrell2012f, bashirzadeh2020confinement, bashirzadeh2021actin, litschel2021protein} to mimic the effects of forces exerted by cell membrane on reconstituted cytoskeletal network. Microfluidic experiments \cite{strelnikova2017direct, kantsler2012fluctuations, manikantan2015buckling, panja2016dynamics, chakrabarti2020trapping, chelakkot2012flow} on semiflexible actin filaments in extensional flows have also shown a stretch-coil transition, with a competition between elasticity and tension causing the actin filaments to buckle. Buckling has also been reported \cite{chen2021compression, ostermeir2010buckling, wada2009semiflexible, cifra2021piston, claessens2006actin} in ring polymers under spherical confinement, which are found in many naturally occurring systems such as viral capsids or circular DNA in bacteria. These non-equilibrium processes may play important roles in many biologically relevant systems, but even the buckling of a single filament in the presence of thermal fluctuations remains poorly understood at equilibrium.

On a macroscopic scale, the buckling of columns and elastic rods has been studied for centuries. According to the Euler buckling theory \cite{kierfeld2008semiflexible}, an elastic rod buckles when the compressive force exceeds a critical value in a sharp transition in the absence of thermal fluctuations. On a microscopic level, buckling of single biomolecules at finite temperature has been observed experimentally \cite{berro2007attachment, murrell2012f, bashirzadeh2021actin} and theoretically \cite{pilyugina2017buckling, ghamari2011buckling}. The buckling transition depends strongly on the mechanical properties of the filament. For example, F-actin has a persistence length ($l_p$) that varies between 10 to 20 $\mu$m \cite{mogilner2005physics, cooper2000structure} and the typical length ($L$) of the filaments found in a filopodium is of the order of 1-10 $\mu$m \cite{mogilner2005physics}. At finite temperatures the filament fluctuates, and the local orientation of the filament vary on the length scale of the persistence length. The mechanical properties of these thermally fluctuating F-actin filaments dictates a cell's response to environmental cues, and so a good understanding of the buckling process in the presence of thermal fluctuations is essential.

To model the buckling process in filamentous biomolecules, the responses to compressive forces are studied in two different ensembles \cite{keller2003relating, giordano2018helmholtz, manca2012elasticity,inequiv}: isometric and isotensional. These ensembles are equivalent to the Helmholtz (fixed volume) and Gibbs (fixed pressure) ensembles of classical statistical mechanics, respectively.  An isometric ensemble is obtained by tethering the end points of the polymer, where the applied force fluctuates. In the isotensional ensemble, the fixed force stretches or compresses the chain and the fluctuating end-to-end distance of a polymer is measured. Classical Euler buckling is in the isotensional ensemble, with a constant compressive force, but in the absence of thermal fluctuations. The equivalence between these two ensembles has been explicitly demonstrated \cite{giordano2018helmholtz, manca2012elasticity} for an unconfined polymer under an applied tension, but differences between the ensembles have been found in the thermodynamic limit in some cases, including confined polymers under tension \cite{inequiv}.  In this paper, we will focus primarily on the statistics of isometric systems. It is important to note that, while isotensional experiments are readily performed using optical tweezers, the most appropriate ensemble {\em{in vivo}} may depend on the system of interest.  For example, the ends of actin filaments inside of the cell can be found trapped in between other biomolecules \cite{blanchoin2014actin, brangwynne2006microtubules, chaudhuri2007reversible, freedman2017versatile}. Stretching or compression of these filaments may be primarily due to the endpoint constraints, rather than a constant applied force.

The wormlike chain (WLC) theory \cite{doi1988theory} is widely used to model semiflexible polymers. The WLC is a continuum theory for slender filaments, incorporating both inextensibility and a length scale, $l_p$, called the persistence length. Deriving statistical quantities from the WLC model exactly involves solving the path integral of a quantum particle \cite{samuel2002elasticity, mehraeen2008end} on the surface of a sphere. A substantial number of numerical studies have explored the process of buckling in single \cite{stuij2019stochastic}, bundled \cite{kierfeld2010modelling, wang2019buckling} and confined \cite{cifra2021piston, hayase2017compressive, evans2013elastocapillary} wormlike chains. While statistical averages can usually be determined numerically using well-established techniques for solving Schr\"odinger equation with a nonlinear potential, it is often the case that analytically tractable results are not easily determined \cite{kierfeld2006buckling, spakowitz2001free, pilyugina2017buckling, bleha2019force, emanuel2007buckling, kurzthaler2018bimodal}. To overcome this limitation, approximate methods such as mean-field theories for wormlike chains \cite{ha1995mean, ha1997semiflexible, morrison2009semiflexible, winkler2003deformation, ghamari2011buckling, tang2015compression, hansen1999buckling}, provide analytically tractable quantitative predictions for a variety of equilibrium statistics that can be more easily applied to experimental data. Previously, Blundell $et \ al.$ \cite{blundell2009buckling} have taken a mean-field theoretical approach to model filament buckling and they have found a weakly first order transition between unbuckled and buckled states in the isotensional ensemble. Despite this theoretical effort, many details of the buckling of a filament, such as the end-to-end distribution functions, remain poorly predicted on a quantitative level. As a result, more work is required to fill the gap in developing a theory for the buckling process in wormlike chains that will have both experimentally accessible predictions as well as biological context.

In this paper, we use an analytically tractable mean field theory combined with Monte-Carlo simulations to determine the statistics of wormlike chains constrained to have fixed end-to-end distance in two and three dimensions.  In Sec. \ref{methods.sec} we describe the mathematical and computational models.  In Sec. \ref{distributions.sec}, we derive the end-to-end distribution for semiflexible chains, recovering a well known result \cite{distributionFunction} in three dimensions but finding poor agreement with simulations in two dimensions.  We show in Sec. \ref{mflp.sec} that the distributions can be brought into agreement with simulations by modifying a parameter in the mean field approximation, and find that the distribution of the transverse position of the midpoint of a chain can be accurately predicted using that modification in Sec. \ref{singleFilSec}.  We find that the distribution transitions from unimodal to bimodal (indicating a continuous phase transition), and that the mean compressive force is nonmonotonic but that the locations of local minima in the force does not coincide with the buckling transition.  Finally, in Sec. \ref{convolve.sec}, we show that the distribution functions predicted in Sec. \ref{singleFilSec} are highly sensitive to small fluctuations in length, and that the theory can be modified to qualitatively reproduce the distribution of transverse positions.  We conclude with a summary of the applicability and limitations of the model.  

\section{Methods}\label{methods.sec}
\subsection{Theory}\label{theory.sec}

The wormlike chain model has been used to describe \cite{teng2021statistical, marantan2018mechanics, everaers2021single, samuel2002elasticity} the statistics and dynamics of a wide range of biomolecules.  This continuum model incorporates two features relevant for a variety of polymers:  inextensibility (fixed length $L$) and a resistance to bending through the Hamiltonian \begin{eqnarray}
\beta H_{wlc}=\frac{l_p}{2}\int_0^L ds (\partial_s\hat\uv)^2\label{truewlc},
\end{eqnarray}
with $\rv$ the position on the polymer, $\hat\uv=|\partial \rv/\partial s|$ the unit vector describing the local direction of the polymer and $\partial/\partial s$ indicates the derivative with respect to the arc-length $s$.  The imposition of an inextensibility constraint makes the WLC model difficult to deal with analytically in all but the simplest cases, and many observables must typically be evaluated numerically \cite{kierfeld2006buckling, pilyugina2017buckling, bleha2019force}.  In many cases \cite{blundell2009buckling, ha1995mean, ha1997semiflexible, morrison2009semiflexible}, analytic progress can be made by relaxing the rigid constraints of inextensibility by constraining the average length of the polymer using a Lagrange multiplier (commonly referred to as a Mean Field (MF) model).   For an unconstrained WLC, an often used MF Hamiltonian is
\begin{eqnarray}
H_0[\uv(s)]=\delta(\uv_0^2+\uv_L^2)+\frac{l_0}{2}\int_0^L ds \dot \uv^2+\lambda\int_0^L ds \uv^2\label{mfham},
\end{eqnarray}
where $l_0$ is a `mean field persistence length' ($l_0\ne l_p$) for the unconstrained WLC, $\lambda$ is a resistance to stretching along the backbone of the chain, and $\uv=\dot\rv$ is the local stretching of the chain in the continuum limit.  For the true WLC, the inextensibility constraint is $|\uv(s)|=1$.   The MF approach chooses the parameters $\lambda$ and $\delta$ such that inextensibility is imposed on average, with $\langle \int_0^L ds\uv^2\rangle_0=L$ and $\langle \uv_0^2+\uv_L^2\rangle_0=1$ (with $\langle \cdots\rangle_0$ a statistical average using the MF Hamiltonian in \ref{mfham}).  The Lagrange multiplier $\lambda$ constrains the length of the chain, while the Lagrange multiplier $\delta$ accounts for the excess fluctuations at the endpoints of the chain \cite{ha1995mean,morrison2009semiflexible,lagrangeElasticity}.  The endpoint fluctuation terms may appear unimportant for long chains, but we will see that deriving an accurate end-to-end distance distribution function requires $\delta$ to be included in the theory (discussed further in Sec. \ref{distributions.sec}).   It is convenient to define the free energy for a wormlike filament as $\Fc_0=-\log\left[\int \Dc[\uv(s)]e^{-\beta H_0}\right]+\lambda L+2\delta$, with the constraints imposed through $\partial\Fc_0/\partial\lambda=\partial\Fc_0/\partial\delta=0$.  The MF Hamiltonian is quadratic, making the calculation of a variety of equilibrium averages straightforward (as discussed in more detail in Sec. \ref{mfsec.sec}).

\subsection{Simulation methodology}

\begin{figure}[htbp]
\begin{center}
\includegraphics[width=.5\textwidth]{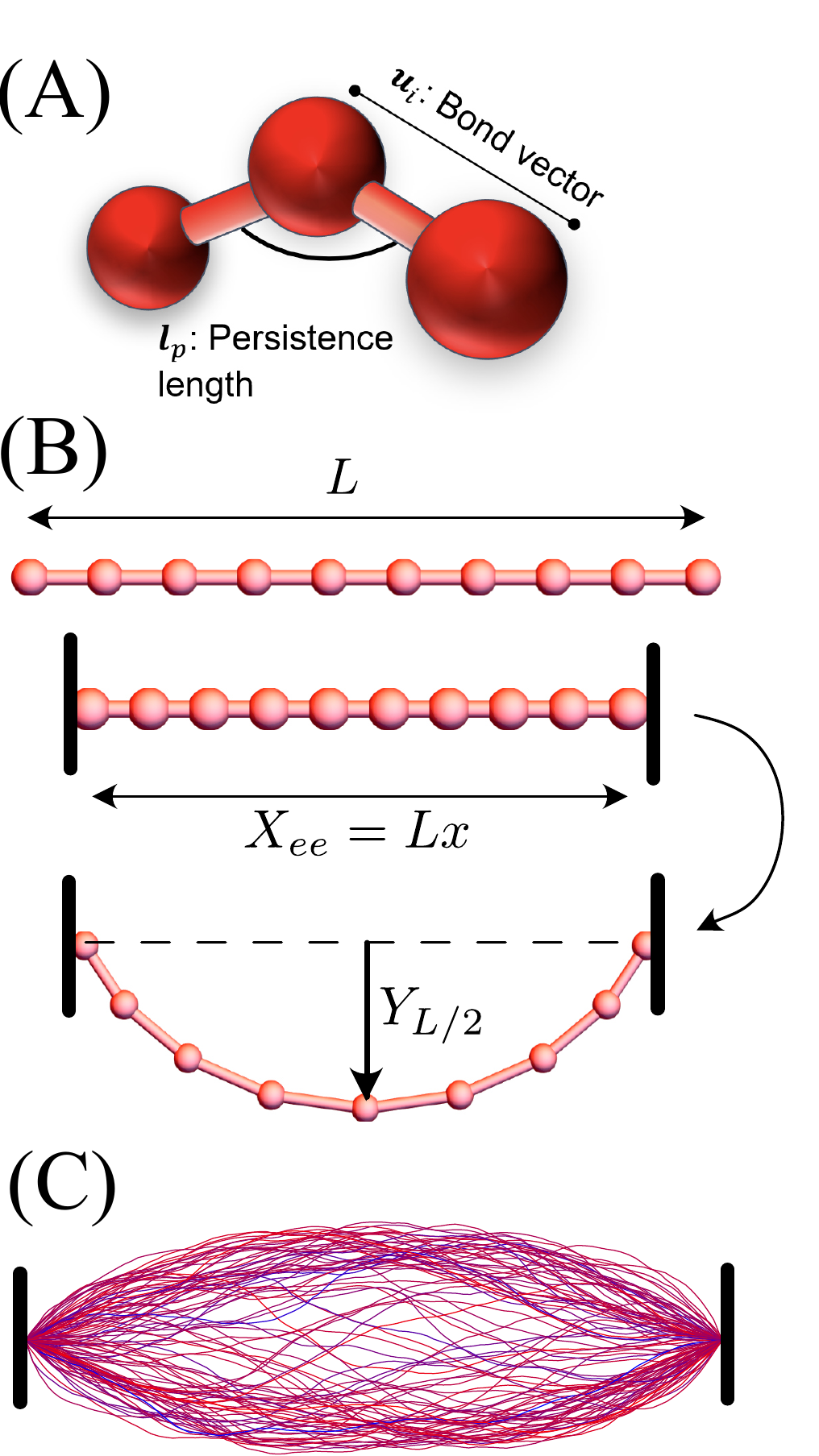}
\caption{Schematic diagram of the WLC model and the buckling process in our simulations.  (A) shows the bending and stretching contributions to the energy in the simulations.  (B) shows the initialization of the simulation:  a chain of length $L$ is compressed uniformly along the backbone to $X_{ee}=0.99L$ and allowed to equilibrate as schematically diagrammed.  The equilibrated configurations from $x=0.99$ are used as the initial conditions for $x=0.98$ after each bond is uniformly compressed by 1\%, and the process is repeated.  (C) shows a sample of 100 equilibrated configurations for the two dimensional simulation with $x=0.94$.  Configurations are colored based on their total bending energy, with redder colors indicating greater energy.  The low density in the center indicates the system has buckled:  the most probable configurations have a nonzero value of $Y_{L/2}=yL$.  }
\label{schematic.fig}
\end{center}
\end{figure}

We simulate a coarse-grained bead-spring chain consisting of $N = 100$ beads with positions indexed as $\rv_i$ and $(N-1)$ bonds with normalized bond vectors given by $\hat{\uv}_i = (\rv_{i+1} - \rv_{i})/|\rv_{i+1} - \rv_{i}|$. The chain length is given by $L = (N-1)a$, where $a$ is the bond length. The Hamiltonian for the discretized WLC consists of two energetic contributions: the bending energy and the stretching energy of the bead-spring chain. In our simulations the bending energy is $U_{\text{bend}} = \kappa\sum_{i} (1 - \hat{\uv}_i \cdot \hat{\uv}_{i+1})$.  To coarse-grain the system, we have used the persistence length \cite{tree2013dna}, $l_p/a = (\kappa - 1 + \kappa \coth\ \kappa)/2(\kappa + 1 - \kappa \coth \ \kappa)$, with $\kappa$ as the bending stiffness parameter. For a stiff chain, $\kappa$ is large ($\kappa \rightarrow \infty$) and so the relation reduces to $l_p/a \approx \kappa - 1/2 + O(e^{-2\kappa})$.  We mapped our coarse-grained simulations onto the typical parameters for actin filaments, with persistence length $l_p = 17 \ \mu$m \cite{gittes1993flexural}.  In most of our simulations, we chose a bending stiffness of $\kappa = 99.5 \ k_{B}T$, so that $l_p\approx L$.  These parameters give a length of approximately $ a \approx 172$ nm between the monomers.   The stretching energy of the harmonic springs connecting the beads is $U_{{s}} =  (\kappa_{s}/2) \sum_{i}(|\uv_i| - a)^2$, where $|\uv_i| = |\rv_{i+1} - \rv_{i}|/a$ is the dimensionless bond length and  $\kappa_{s}a^2 = 490  \ k_{B}T$ is the stretching stiffness used in our simulations.

 We first perform Monte-Carlo simulations of a WLC in the absence of endpoint pinning or compressive forces. We initialize each chain as a rod of length L, and generate trial configurations by moving a randomly chosen monomer in a random direction sampled from a normal distribution with a mean zero and standard deviation $0.25 \ a$. The energy difference ($\Delta E$) between the modified configuration and the previous configuration is calculated to check if the trial move is accepted or rejected. The configurations are accepted following the Metropolis criterion \cite{frenkel2001understanding} which says the probability of acceptance is directly proportional to the Boltzmann factor, $p_{\text{accept}} = e^{-\beta \Delta E}$ with $\beta = 1/k_{B}T$. There are $10^9$ MC steps in between the initially grown chain and the final equilibrated chain configurations which is the data used for plotting the end-to-end distance probabilities. A total of 1600 configurations are produced after equilibration.

To obtain the buckling statistics in the isometric ensemble, we pin the chain at the end-points in the x-direction so that the end-to-end distance is fixed (with $X_{ee} = x L$). For $X_{ee}\approx L$, the mean force applied to the filament is positive, stretching the filament in the $x-$direction.  For sufficiently low $X_{ee}$, the mean force will become negative, and the chain will be compressed (and we expect the filament will buckle for some critical $X_{ee}$).   We use the unpinned simulation method as described above to perform these simulations, modified to keep the end-to-end distance fixed by never selecting the endpoints as the monomers that are moved ($\rv_0$ and $\rv_N$ are unchanged in all trial moves).  We initialize the simulation with a rod of length $X_{ee}=0.99 L$ by compressing each bond vector's length by 1\%.  We equilibrated 5$\times 10^4$ configurations in two dimensions and $\approx 8400$ configurations in three dimensions by performing $\approx 10^9$ MC steps per simulation.  After equilibration, these configurations are compressed by 1\% by reducing the length of each bond uniformly and equilibrating again.  This procedure was followed reducing $x$ by 0.01 at every iteration, to $x = 0.94$ for two dimensions and $x = 0.87$ for three dimensions.

\section{Results}

\subsection{\label{mfsec.sec}The mean field approach for wormlike chains}

The MF method replaces rigid local constraints with averaged global constraints (as described in Sec. \ref{theory.sec}). A difficulty with the mean field approach is determining the relationship between $l_0$ (the MF persistence length in eq. \ref{mfham}) and $l_p$ (the true persistence length of the polymer in eq. \ref{truewlc}).  With {\em{a priori}} knowledge of the exact mean squared curvature of the WLC, it is straightforward to use $l_0$ as an additional Lagrange multiplier\cite{lagrangeElasticity}, choosing the optimal value of $l_0$ by constraining $\left\langle\int_0^L ds\dot\uv^2\right\rangle_0$ to a known value.  In this paper, we will generally not have this knowledge {\em{a priori}}, and must find an approximate method for identifying a relationship between $l_0$ and $l_p$.  In ref \cite{distributionFunction}, the connection is made by recognizing that an {\em{unconstrained}} WLC has $\langle \uv_0\cdot\uv_L\rangle_{0}=e^{-3L/2l_0}$ in three dimensions, in comparison to the exact result $\langle \uv_0\cdot\uv_L\rangle_{exact,3d}=e^{-L/l_p}$ (with $\langle \cdots\rangle_{exact,3d}$ the true statistical average in three dimensions).  This suggests the replacement $l_0=3l_p/2$ in 3 dimensions.    A similar calculation in two dimensions shows that $\langle \uv_0\cdot\uv_L\rangle_{0}=e^{-L/l_0}$, in comparison to the exact $\langle\uv_0\cdot\uv_L\rangle_{exact,2d}=e^{-L/2l_p}$, suggesting the substitution $l_0=2l_p$ in two dimensions.  This substitution $l_0=3l_p/2$ in three dimensions has been used in multiple contexts \cite{ha1995mean,ha1997semiflexible,distributionFunction,morrison2009semiflexible}; we are not aware of this MF formalism utilized in two dimensions.

\subsection{Distribution functions in two and three dimensions}
\label{distributions.sec}

An accurate derivation of the end-to-end distance distribution for a WLC using the MF theory has been previously accomplished in three dimensions.  To our knowledge an equivalent two-dimensional distribution derived using this MF approach has not been reported in the literature, and we will compute it in this section.  To determine the distribution function, we will find it convenient to define the Hamiltonian $H_1=\frac{1}{2}l_1\int_0^L ds \dot\uv^2+\lambda \int_0^L ds \uv^2+\delta(\uv_0^2+\uv_L^2)$, which is identical to the definition of $H_0$ in eq. \ref{mfham} but with a mean field persistence length $l_1$ (with $l_1\ne l_0$ in general).   The distribution function in two and three dimensions are determined by computing the constrained free energy $e^{-\Fc(\Rv_{ee})}=\int \D[\uv(s)]e^{- H_1+\lambda L+2\delta} \delta\left(\Rv_{ee}-\int_0^L ds\uv(s)\right)=\left\langle e^{\lambda L+2\delta}\delta(\Rv_{ee}-\int_0^L ds\uv(s))\right\rangle_1$, with $\langle \cdots\rangle_1$ an average with respect to $H_1$.  The constraint of inextensibility is imposed on average by requiring $\partial\log[\Fc(\Rv_{ee})]/\partial\lambda=\partial\log[\Fc(\Rv_{ee})]/\partial\delta=0$. We can readily evaluate this via $e^{-\Fc(\Rv_{ee})}=e^{\lambda L+2\delta}\int d^d \qv e^{i\qv\cdot\Rv_{ee}}\left\langle\exp\left(i\qv\cdot\int_0^L ds\uv(s)\right)\right\rangle_1$.  This calculation was performed in ref \cite{distributionFunction} in three dimensions and is straightforward in two dimensions by completing the square in the Hamiltonian, with $\lambda\int_0^L ds (\uv^2-i\qv\cdot\uv)=\qv^2 L/2\lambda+\lambda \int_0^L ds \vv^2$ for $\vv=\uv-i\qv/2\lambda$.  The path integrals can be evaluated using the propagator for the quantum harmonic oscillator in $d$-dimensions, with $Z(\uv_0,\uv_L;L)=(l_1\Omega/2\pi\sinh(\Omega L))^{d/2}\,\exp[-\frac{l_1\Omega}{2}(\uv_0^2+\uv_L^2)\coth(\Omega L) + l_1\Omega\uv_0\cdot\uv_L\csch(\Omega L)]$ and with $\Omega=\sqrt{2\lambda/l_1}$.  The free-energy minimizing equations for $\delta$ and $\lambda$ are unwieldy, but are simplified considerably if we assume $L$ is large and $\Rv_{ee}^2=r^2L^2$.  Retaining terms of order $L$ and replacing $\sinh(\Omega L)\approx \cosh(\Omega L)\approx e^{\Omega L}/2$ yields  
\begin{eqnarray}
\frac{2d}{2\delta+l_1\O}=2\qquad \Omega l_1(1-r^2)=\frac{d}{2}
\end{eqnarray}
in $d=2$ or 3 dimensions, with solutions $\Omega=d/2l_1(1-r^2)$ and $\delta=d(1-2r^2)/4(1-r^2)$.  These values are substituted into the free energy $\Fc(r)=\frac{1}{d}\left(\frac{\pi^{1/2}l_1^2\O^{3/2}}{L^d}\right)^d\exp\left(-\frac{Ll_1\O^2}{2}(1-r^2)+\frac{dL\O}{2}+2\delta\right)$.  We must normalize these distributions so that $L^d\int_0^1d^d\rv e^{-\Fc(\rv)}=1$.  The integrals are straightforward to evaluate with the substitution $u=1/(1-r^2)$, and $\rv=\Rv_{ee}/L$ we find to leading order in $L/l_p$,
\begin{eqnarray}
P_{2d}(\rv)&=&\frac{L^2e^{L/2l_1}}{2\pi l_1L(L+2l_1)}\ \frac{e^{-L/2l_1(1-|\rv|^2)}}{(1-|\rv|^2)^{3}}\label{prob2d},\\
P_{3d}(\rv)&=& \frac{3^{6}L^{7/2}e^{9L/8l_1}}{2^{9/2}\pi^{3/2}(27 L^2+72 Ll_1+80 l_1^2)}\ \frac{e^{9L/(8l_1(1-|\rv|^2)}}{(1-|\rv|^2)^{9/2}}\label{prob3d}.
\end{eqnarray}
Here we neglect higher order contributions in $L/l_1$, but we will find below that this leading order approximation is surprisingly accurate. Note that equations 4 and 5 are vector distributions; when computing the distribution of the magnitude of the end-to-end distance, the volume elements $2\pi r$ or $4\pi r^2$ for two and three dimensions respectively, should be included.   Terms in the three dimensional distribution differs from that in Ref. \cite{distributionFunction} because we have not replaced the mean field persistence length $l_1$ in terms of the true persistence length $l_p$ in eq. \ref{prob3d} yet (discussed further below).   The factor of $\delta$ plays a critical role in computing the distribution function by suppressing the excess fluctuations at the endpoints, as the end-to-end distribution function would be $\propto (1-|\rv|^2)^{d/2}$ with $\delta=0$ rather than $\propto (1-|\rv|^2)^{3d/2}$ we have found here.   

\subsection{The mean-field persistence length in two and three dimensions\label{mflp.sec}}

\begin{figure}[htbp]
\begin{center}
\includegraphics[width=\textwidth]{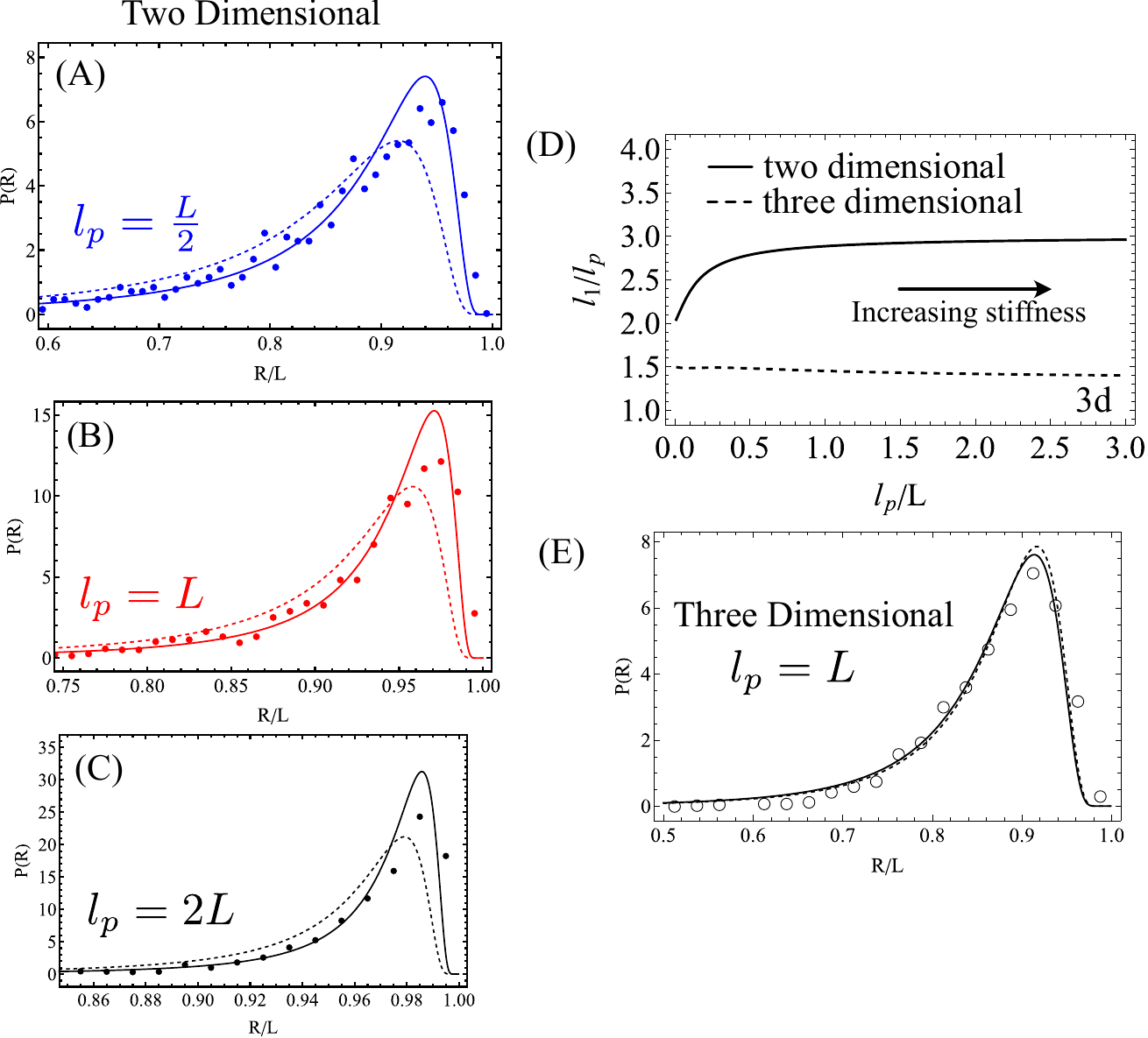}

\caption{(A-C)  End-to-end distribution functions for 2 dimensional WLCs at the indicated persistence length.  The dashed lines correspond to eq. \ref{prob2d}, with the substitution $l_1=2l_p$, while the solid lines correspond to the substitution in eq. \ref{lp2d} (with from left to right $l_1\approx 2.78l_p$, $l_1\approx 2.92l_p$, and $l_1\approx 2.96l_p$).  A significant improvement in the agreement is observed using eq. \ref{lp2d}, consistent with Table. \ref{agree2d.tab}.  (D)  The ratio $l_1/l_p$ in two (solid black line) and three (dashed black line) dimensions.  In three dimensions, $l_1/l_p$ is weakly varying between 1.35 to 1.50.  In two dimensions, the change is more significant (varying between 2 and 3), leading to the differences seen in the dashed lines in (A-C).  (E)  End-to-end distribution functions for $l_p=L$ in three dimensions.  The dashed lines correspond to eq. \ref{prob3d} with the substitution $l_1=\frac{3}{2}l_p$, while the solid lines correspond to the the choice of $l_1$ such that $\langle \Rv_{ee}^2\rangle_{MF,3}=\langle\Rv_{ee}^2\rangle_{exact,3}$.    In 3 dimensions, the distributions are nearly identical using either definition for $l_1$, suggesting the mean field theory does not depend sensitively on the method by which $l_1$ is chosen in three dimensions, consistent with (D).}
\label{goodbadfit.fig}
\end{center}
\end{figure}

As discussed in the previous section, the mean field persistence length $l_1$ is not identical to the true persistence length of the chain, and without knowing an exact $\langle \int_0^L ds\dot\uv^2\rangle$ we cannot variationally determine $l_1$ explicitly.  In Ref\cite{distributionFunction}, it is argued that we should expect $l_1=l_0=3l_p/2$ for $d=3$, based on the correlation function $\langle \uv_0\cdot\uv_L\rangle_{d=3}=e^{-3 L/2l_0}$.  End-to-end distributions for three dimensional WLCs have been accurately recovered using eq. \ref{prob3d} using this approximation.  In two dimensions, a similar argument would suggest $l_1$ should be replaced with $l_0=2l_p$ for $d=2$. Surprisingly, we find a fairly poor agreement with the simulated end-to-end distance distribution functions of a 2 dimensional WLC (the dashed lines in Fig. \ref{goodbadfit.fig}(A-C)):  neither the mean nor the most probable end-to-end distances agree well with the simulated data with the substitution $l_1=2l_p$.  The mean squared end to end distance is significantly underestimated in comparison to the simulations (see Table \ref{agree2d.tab}).   The most probable value $r^{max}$ is also underestimated, and the probability of finding $r=r^{max}$ is also low compared to the simulations.  This suggests that the distributions are not well predicted using the MF method with the substitution $l_1=2l_p$. 

   \begin{table}
\centering
\caption{Comparison of the simulated and theoretical distributions using different values of $l_1$:  either $l_1=2l_p$ (equal to $l_0$, predicted by the decay in the correlation function as described in Sec. \ref{distributions.sec}), or defining $l_1$ using eq. \ref{lp2d}.  We compute the ratio of the simulated vs MF mean squared end-to-end distances and the Kullback-Liebler divergence, $D$.  Using eq. \ref{lp2d} gives a much lower ratio of the simulated vs theoretical end-to-end distances, and reduces $D$. }
\label{agree2d.tab}
\begin{tabular}{|*{7}{c|}}  
\hline
 & \multicolumn{3}{c|}{$\langle R_{ee}^2\rangle_{sim}/\langle R_{ee}^2\rangle_{MF}$} 
 & \multicolumn{3}{c|}{K-L Divergence $D$ }  \\
\cline{2-7}
  Quantity      & $l_p=\frac{L}{2}$ &   $l_p={L}$    & $l_p=2{L}$   &           $l_p=\frac{L}{2}$ &   $l_p={L}$    & $l_p=2{L}$\\
  \hline
 using $l_1=2l_p$ &  1.11 & 1.07 & 1.04 & 33.2 & 22.3 & 22.7 \\
 using eq. \ref{lp2d} & 1.01 &  1.00 &  1.01& 23.1 & 7.34 & 2.86\\     
    \hline
    \end{tabular}
\end{table}

In order to improve the agreement, we instead choose $l_1$ such that $\langle \rv^2\rangle_{true,d}=\int d^d\rv P(\rv)$, where $\langle \rv^2\rangle_{true,d}$ is the exact mean squared end-to-end distance for a WLC in $d$ dimensions:  $\langle \Rv_{ee}^2\rangle_{exact,3d}=2Ll_p+2l_p^2(e^{-L/l_p}-1)$, and $\langle \Rv_{ee}^2\rangle_{exact,2d}=4Ll_p+2l_p^2(e^{-L/2l_p}-1)$.  These can be compared to the mean-squared end-to-end distances predicted by the MF theory in terms of $l$, with 
\begin{eqnarray}
\langle \Rv_{ee}^2\rangle_{MF,2}&=&\int_0^1 dr\ 2\pi r^3 P_{2d}(r)=\frac{2L^2l_1}{L+2l_1},\\
\langle \Rv_{ee}^2\rangle_{MF,3}&=&\int_0^1 dr\ 4\pi r^4 P_{3d}(r)=\frac{4L^2 l_1(9L+20l_1)}{27 L^2+72 Ll_1+80 l_1^2}.
\end{eqnarray}
In two dimensions, equating the exact and and MF predictions for the mean squared end-to-end distance yields 
\begin{eqnarray}
l_1=l_p\frac{2(1-2(1-e^{-L/2l_p})\frac{l_p}{L})}{1-4\frac{l_p}{L}+8(1-e^{-L/2l_p})\frac{l_p^2}{L^2}}\qquad \mbox{(2 dimensions)}\label{lp2d}.
\end{eqnarray}
For flexible chains with $L\gg l_p$, we find eq. \ref{lp2d} reduces to $l_1\approx 2l_p$, recovering the substitution for the bending correlation functions to recover their unconstrained expected values.  However, in the limit of $l_p\gg L$, eq. \ref{lp2d} reduces to $l_1\approx 3l_p-L/8+O(l_p^{-1})$.  This change in the coefficient (from 2 to 3) suggests $l_1/l_p$ may vary significantly with $r$ for $l_p\gtrsim L$, and that the substitution $l_1=2l_p$ would only be applicable in the limit of $l_p\to 0$.   In Fig. \ref{goodbadfit.fig}(D), the solid line shows eq. \ref{lp2d} as a function of $l_p/L$, and we see that significant deviations from $l_1/l_p=2$ occur even for moderate stiffnesses.  Our method ensures the mean of the theoretical distribution matches the exact value, and unsurprisingly this reduces the difference in the mean squared end-to-end distance.  We also compare the Kullback-Leibler (KL) divergence, $D=\sum_i P_{sim}(r_i)\log\left[P_{sim}(r_i)/P_{MF}(r_i)\right]$ between the simulated and theoretical distributions, and find $D$ is significantly reduced when eq. \ref{lp2d} is used to relate $l_1$ to $l_p$.  

In three dimensions, solving for $\langle \Rv_{ee}^2\rangle_{MF,3}=\langle\Rv_{ee}^2\rangle_{exact,3}$ leads to a more complicated expression for the relationship between $l_1$ and $l_p$, which is easily evaluated numerically.  We find $1.35\le l_1/l_p\le 1.5$ in three dimensions, shown in Fig. \ref{goodbadfit.fig}(D).  The numerical differences for stiff and flexible chains are much small in three dimension than in two dimension and suggests that using $l_1\approx 3l_p/2$ for the MF persistence length may be approximately valid even for stiff chains (consistent with ref \cite{distributionFunction}).  The agreement in three dimensions is shown in Fig \ref{goodbadfit.fig}(E) for $l_p/L=1$ and agrees well.

\subsection{Buckling of a compressed filament\label{singleFilSec}}

To estimate the end-to-end distance at which the chain will buckle, we determine the distribution function for the transverse position of a filament pinned at both endpoints, with the expectation that the emergence of a bimodal distribution indicates a buckling transition (sketched in Fig. \ref{schematic.fig}(C)).   In two and three dimensions, we assume the end-to-end distance to be in $x$-direction with magnitude $X_{ee}$,  and wish to compute the $y$-position along the backbone at $s$ ($Y_s$).  In two dimensions, this distribution can be determined by computing 
\begin{eqnarray}
P(Y_{s};X_{ee})=\bigg\langle \delta\bigg(X_{ee}-\int_0^L ds u_x\bigg)\delta\bigg(\int_0^L ds u_y\bigg)\delta\bigg(Y_s-\int_0^s ds' u_y(s')\bigg)\bigg\rangle_1\label{distro2dinit},
\end{eqnarray}
where $\uv=\dot\rv=(u_x,u_y)$ is the tangent vector.   In three dimensions a similar expression can be developed including an additional term $\delta\left(\int_0^L ds u_z(s)\right)$ in the average.  Note that in two and three dimensions, we do not impose a constraint on the direction of the pinned bonds (that is, $\uv(0)$ and $\uv(L)$ are not constrained to be aligned with the $x$-axis).  This assumption is discussed in more detail in the conclusions.  This average is readily evaluated using the Fourier Transform
\begin{eqnarray}
P(Y_{s};X_{ee})&=&\int\frac{d^d\kv dq}{(2\pi)^{d+1}}e^{-i\kv\cdot\Rv_{ee}-iqY_s}\exp \bigg(i\kv\cdot\int_0^Lds' \uv(s')+iq\int_0^s ds'u_y(s')\bigg)e^{-\beta H_1},
\end{eqnarray}
with $\Rv_{ee}=(X_{ee},0)$ in two dimensions and $(X_{ee},0,0)$ in three dimensions. Because the MF Hamiltonian is quadratic, the integral can be evaluated in a straightforward fashion.  As was the case for the end-to-end distribution function, it is extremely useful to take the limit of $L\gg 0$ so that the hyperbolic trigonometric functions can be replaced with exponentials.  We further assume that $e^{s\Omega/2}\gg e^{-s\Omega/2}$, meaning the point $s$ is far from the endpoints.  In this limit, we find after some algebra that in $d=2$ or 3 dimensions the distribution function is
\begin{eqnarray}
P(Y_s;X_{ee})\propto \frac{\Omega^{3d/2}e^{2\delta-dL\Omega/2+l_1L\Omega^2/2}}{(2\delta+l_1\Omega)^dg^{1/2}(\Omega)}\exp\bigg(-\frac{l_1 X_{ee}^2 \Omega^2}{2L}+\frac{Ll_1 Y_s^2\Omega^2}{g(\Omega)}\bigg)\label{bucklingDist0},
\end{eqnarray}
with $g(\Omega)=1-2s(1-\frac{s}{L})\Omega$.  It is convenient to write $X_{ee}=x L$, $Y_s=y L$, and $s=\sigma L$.  Taking the limit of $L\to\infty$ imposing the inextensiblity constraints on average using $\partial \log(P)/\partial\delta=\partial\log(P)/\partial\lambda=0$ gives the mean field solutions
\begin{eqnarray}
\delta=\frac{d}{2}-\frac{l_1\Omega}{2}\qquad\qquad\Omega=\frac{d}{2l_1}\bigg(1-x^2-\frac{y^2}{\sigma(1-\sigma)}\bigg)^{-1}\label{bucklingRoots}.
\end{eqnarray}
Substitution of eq. \ref{bucklingRoots} into eq. \ref{bucklingDist0} gives the distributions to leading order in $L$
\begin{eqnarray}
P(y;x)\propto \bigg(1-x^2-\frac{y^2}{\sigma(1-\sigma)}\bigg)^{-3d/2-1}\exp\bigg[\frac{(2d-3)L}{2l_1}\bigg(1-x^2-\frac{y^2}{\sigma(1-\sigma)}\bigg)^{-1}\bigg]\label{distribution}.
\end{eqnarray}
These distributions cannot be integrated analytically in terms of elementary functions (unlike the case of the end-to-end distribution function in eq. \ref{prob2d} or \ref{prob3d}), and we will numerically evaluate the normalization factor. 

\begin{figure}[htbp]
\begin{center}
\includegraphics[width=.6\textwidth]{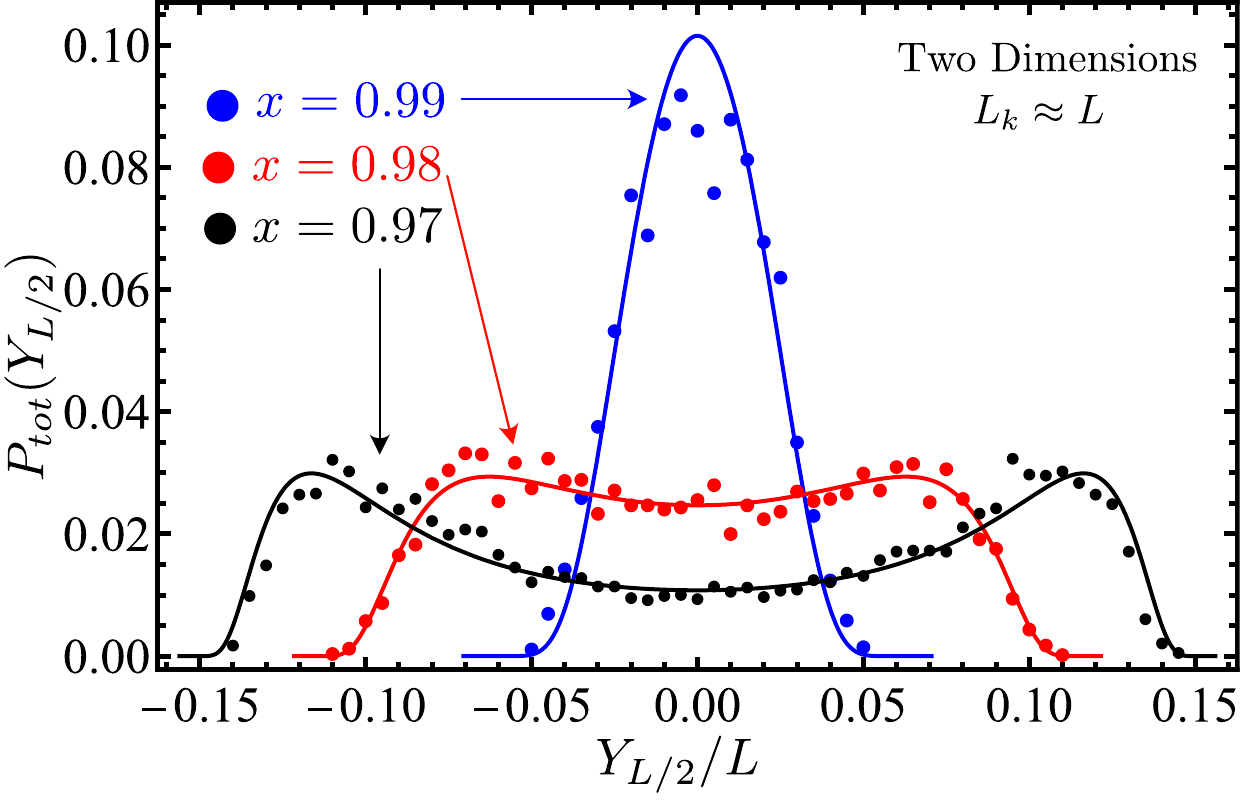}
\caption{Transverse position distribution of the mid-point of a two-dimensional filament with $l_p=L$.  Points represent the $y$ component of the midpoint of the simulated data for which the length of the $k^{th}$ simulated chain is $L_k\approx (N-1)a$ (permitting variations of around 0.1\%), and lines correspond to the predictions of eq. \ref{distribution}, replacing $l_1$ with eq. \ref{lp2d}.  Note that there are no free parameters in the fitting.  Blue refers to $X_{ee}/L=0.99$, red 0.97, and black 0.95.   }
\label{2dfit.fig}
\end{center}
\end{figure}

\begin{figure}[htbp]
\begin{center}
\includegraphics[width=.75\textwidth]{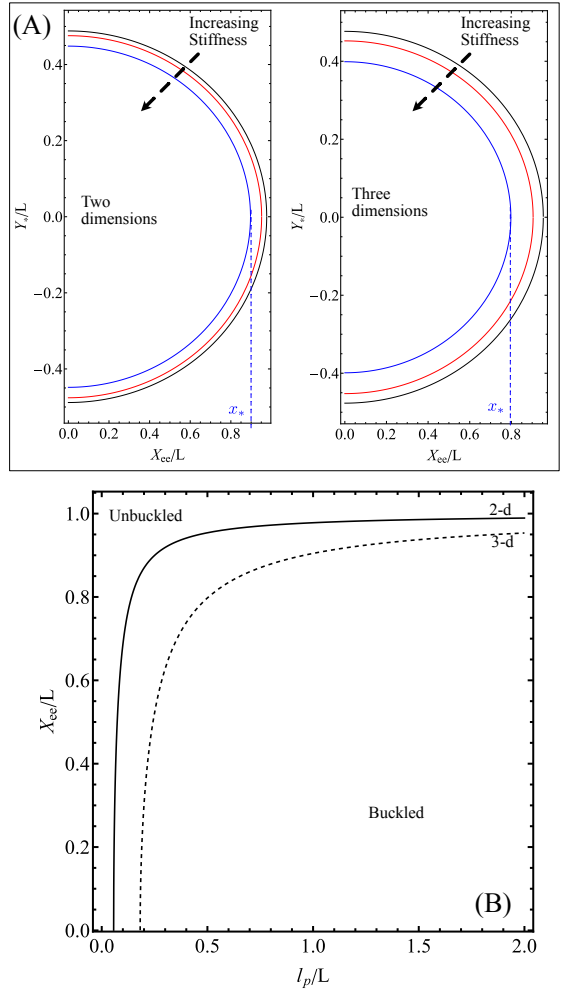}
\caption{\label{phase.fig}(A) The most probable value of the transverse position of the midpoint of the polymer in two and three dimensions, with $l_p=L/2$ (blue), $L$ (red), and $2L$ (black). Buckling occurs when the most probable value of the transverse position is non-zero (when $x=x_*=\sqrt{1-c_dL/l_1}$).  (B) Phase diagram of the transition between unbuckled and buckled filaments.  The transition occurs for more flexible chains in two dimensions than three dimensions, and in either dimensionality there is a minimum persistence length required for buckling to occur.}
\end{center}
\end{figure}

The length of the chain fluctuates in the simulations, with the length of the $k^{th}$ filament defined as $L_k=\sum_i |\rv_i^{(k)}-\rv_{i-1}^{(k)}|$, while eq. \ref{distribution} assumes that the filament has a fixed length $L$.  In order to compare the theory to simulations, in Fig. \ref{2dfit.fig} we show the distribution of $y=Y_{L/2}/L$ for only data with $0.999<L_k/L<1.001$ (selecting only $\approx 10\%$ of our simulated configurations with an approximately fixed length; length fluctuations will be accounted for in Sec. \ref{convolve.sec}).  We plot the distribution in Eq. \ref{distribution} with $l_1$ defined in Eq. \ref{lp2d}, and see that the distribution is well predicted by the theory for (almost) constant L.  When $x=X_{ee}/L\approx 1$, the distributions are unimodal and peaked around $y=0$, while bimodal distributions are observed when $x$ is sufficiently small.  These bimodal distributions indicate the filament has buckled: the most probable location is not centered at $Y_{L/2} = 0$, but instead at a finite value.  The distribution functions have critical points as a function of $y$ at $y_*=0$ or $y_*=\pm \left(\frac{1-x^2}{4}-\frac{c_d L}{4l_1}\right)^{1/2}$, with $c_d=1/8$ in two dimensions and $c_d=3/11$ in three dimensions.  The nonzero peaks become real when $x<x_*=\left(1-\frac{c_dL}{l_1}\right)^{1/2}$ with $x_*$ the critical compressed distance at which buckling occurs.  Buckling is a continuous phase transition as shown in Fig. \ref{phase.fig}(a), with $y_*=0$ when $x\ge x_*$, consistent with the observations in Ref. \cite{pilyugina2017buckling}.  The phase diagram of the buckling is shown in Fig. \ref{phase.fig}(b) for both two and three dimensions.  The onset of buckling occurs for larger $x_*$ in two dimensions than in three dimensions; for our parameters of $l_p\approx L$ the buckling occurs at $x_*\approx 0.978$ in two dimensions and $x_*=\frac{3}{\sqrt{11}}\approx 0.905$ in three dimensions.  Note that we also predict that buckling {{will not occur}} for sufficiently flexible chains (when $l_1/L\le c_d$) since $x_*$ is imaginary for low $l_1$.  In terms of the true persistence length, this corresponds to $\frac{l_p}{L}\approx 0.056=\alpha_2$ in two dimensions and $\frac{l_p}{L}=\frac{2}{11}\approx 0.182=\alpha_3$ in three dimensions.  

\begin{figure}[htbp]
\begin{center}
\includegraphics[width=0.8\textwidth]{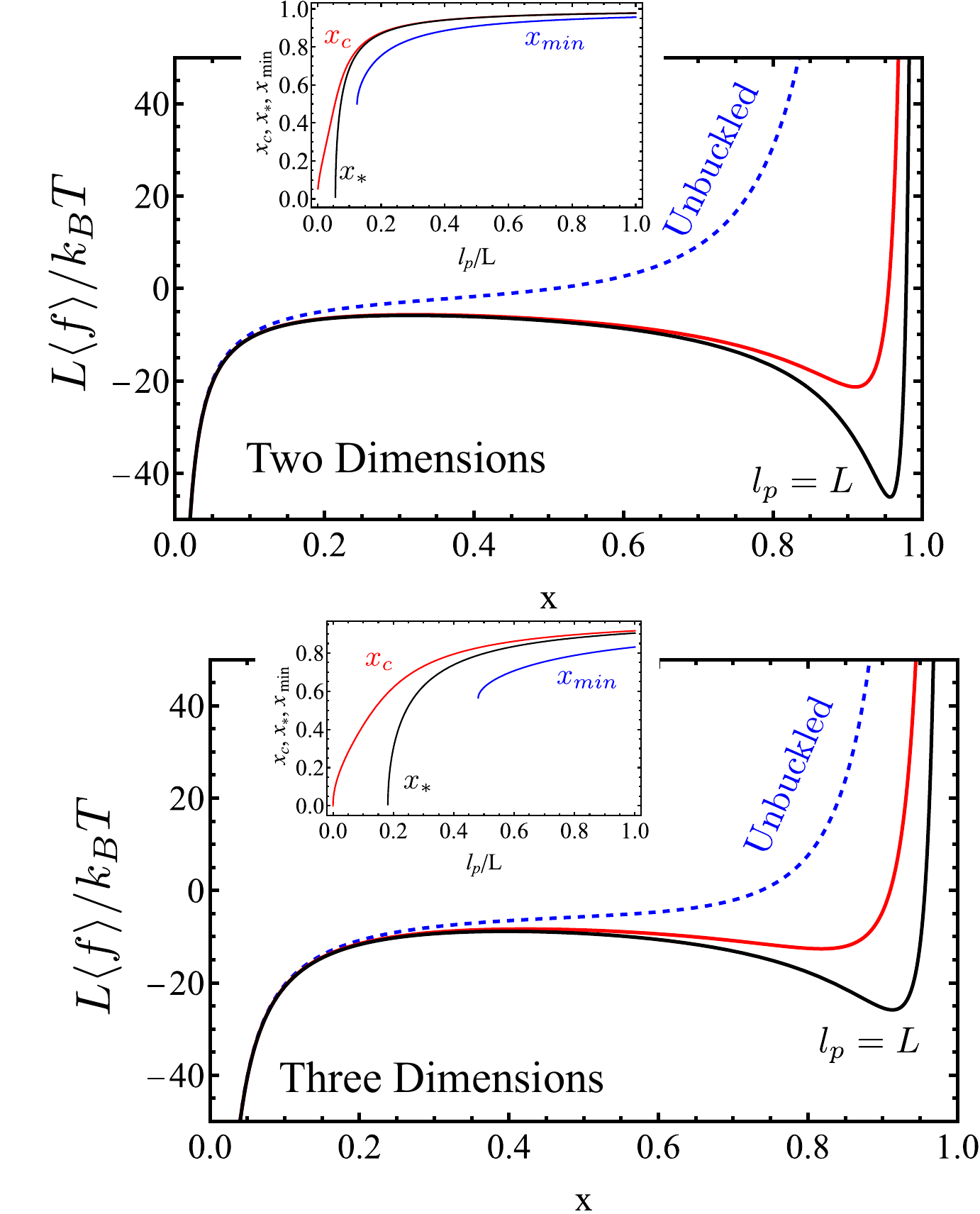}
\caption{Mean force on the filaments in two (A) and three (B) dimensions as a function of the fixed extension $x=X_{ee}/L$.  Shown are $l_p/L=1$ (black) and 1/2 (red), as well as the stiffnesses $\alpha_d$ at which buckling does not occur (dashed blue line). The insets for each show the extension at which the force becomes compressive ($x_c$), the extension at which the phase transition occurs ($x_*$), and the location of the local minimum of the force ($x_{min}$) as a function of stiffness.  Buckling does not coincide with the minimum in the compressive forces, and in two dimensions the onset of buckling is very close to the onset of compression even for fairly flexible chains. }
\label{force.fig}
\end{center}
\end{figure}

Pinning the endpoints of a wormlike chain fixes the end-to-end distance, and thus the compressive force fluctuates.  This is distinct from the problem of Euler buckling, which predicts a first-order phase transition in the $T\to 0$ limit.  We can readily compute the mean compressive force via $L\beta \langle f_c\rangle=-\partial\log[P(r)]/\partial r$, and find 
\begin{eqnarray}
L\beta \langle f_c(r)\rangle &=&-\frac{8x}{(1-x^2)}+\frac{ Lx}{l_1(1-x^2)^2}+\frac{1}{x} \quad\mbox{2 dimensions}\label{f2d},\\
&=&-\frac{(1-9x^2)}{x(1-x^2)}+\frac{ Lx}{l_1(1-x^2)^2}+\frac{2}{x}\quad\mbox{3 dimensions}\label{f3d}.
\end{eqnarray}
These forces are shown in Fig. \ref{force.fig}, with $r = x$ and the substitution of eq. \ref{lp2d} for $l_1$ in two dimensions (Fig. \ref{force.fig}(A)) and $l_1=3l_p/2$ in three dimensions (Fig. \ref{force.fig}(B)). As expected, for $x \approx 1$ the mean forces are positive and elongate the chain. The onset of the compressive forces are earlier in two dimensions than in three, at a higher compression ratio $x$ (as in Fig. \ref{phase.fig}(b)).  Substitution of $x=x_*$ into eq. \ref{f2d}-\ref{f3d} give a mean compressive force $\langle f_c(r_*)\rangle=-\frac{33k_BT l_p}{8L^2}\sqrt{1-\frac{2L}{11l_p}}$ in three dimensions.  The standard Euler buckling result at $T=0$ predicts a critical force at \cite{pilyugina2017buckling, blundell2009buckling} $f_E=-\pi^2\frac{k_BT l_p}{L^2}$.  Our theoretical prediction thus gives the same scaling of the mean compressive force with $l_p$ and $L$ as has been used previously.  The scaling coefficient is about 60\% lower than the standard Euler buckling coefficient in the limit of $l_p\to\infty$.  Ref \cite{pilyugina2017buckling} finds that the critical buckling force on a wormlike chain falls somewhere between $0.67 f_E$ and $1.33 f_E$ in the presence of thermal fluctuations and a constant compressive force. This apparent inconsistency is due to the fact that we use a fixed end-to-end distance (Helmholtz) ensemble, rather than a fixed force (Gibbs) ensemble.  For stiff chains in two dimensions, the buckling transition nearly coincides with the onset of a mean compressive force. It is also worth noting that the scaling coefficients in the mean field approach are expected to be correct within an order of magnitude. A surprising result in two dimensions is that we find $f_c(x_*)=k_BT/L\sqrt{1-c_d L/l_1}$, with the scaling coefficient of the $l_p/L^2$ term vanishing.  We thus do not recover the scaling for the Euler buckling in two dimension, but rather find that the onset of of the buckling transition occurs when the mean compressive force is $\approx -k_BT/L$ for stiff chains (discussed below).  This again may be due to inaccuracies in the scaling coefficient (if $c_2\ne 1/8$, the usual Euler buckling scaling re-emerges with a different scaling coefficient).  

In the insets of Fig. \ref{force.fig}, we show the extension at which the forces become compressive ($x_c$, the compression at which \ref{f2d} or \ref{f3d} become negative), the onset of buckling at $x_*$, and the location of the local minimum of the force ($x_{min}$).  We see the buckling transition occurs between the onset of a compression and the local minimum, but that for stiff chains $x_*\approx x_c$, this means the chains of even moderate stiffness buckle almost immediately in two dimensions when a mean compressive force is applied.  In three dimensions, the buckling transition is between the onset of compression and the local minimum, but $x_*\not\approx x_c$ unless the chains are very stiff.  In both two and three dimensions, the local minima in the compressive forces do not coincide with the onset of the buckling transition.   Much like the absence of the buckling for sufficiently flexible chains, we see these local minima in the compressive forces cease to exist for sufficiently low $l_p/L$.    It is interesting to note that the scaling coefficient of the Euler buckling solution is independent of the dimensionality\cite{pilyugina2017buckling} ($f_E=-\pi^2k_BTl_p/L^2$ in two and three dimensions), since the action-minimizing $T=0$ solution has a constant azimuthal angle.  Here, we see that the dimensionality does affect the scaling coefficient of the compressive force at finite $T$ with fixed endpoints (rather than constant force).

\subsection{Accounting for fluctuations in length\label{convolve.sec}}

\begin{figure}[htbp]
\begin{center}
\includegraphics[width=.9\textwidth]{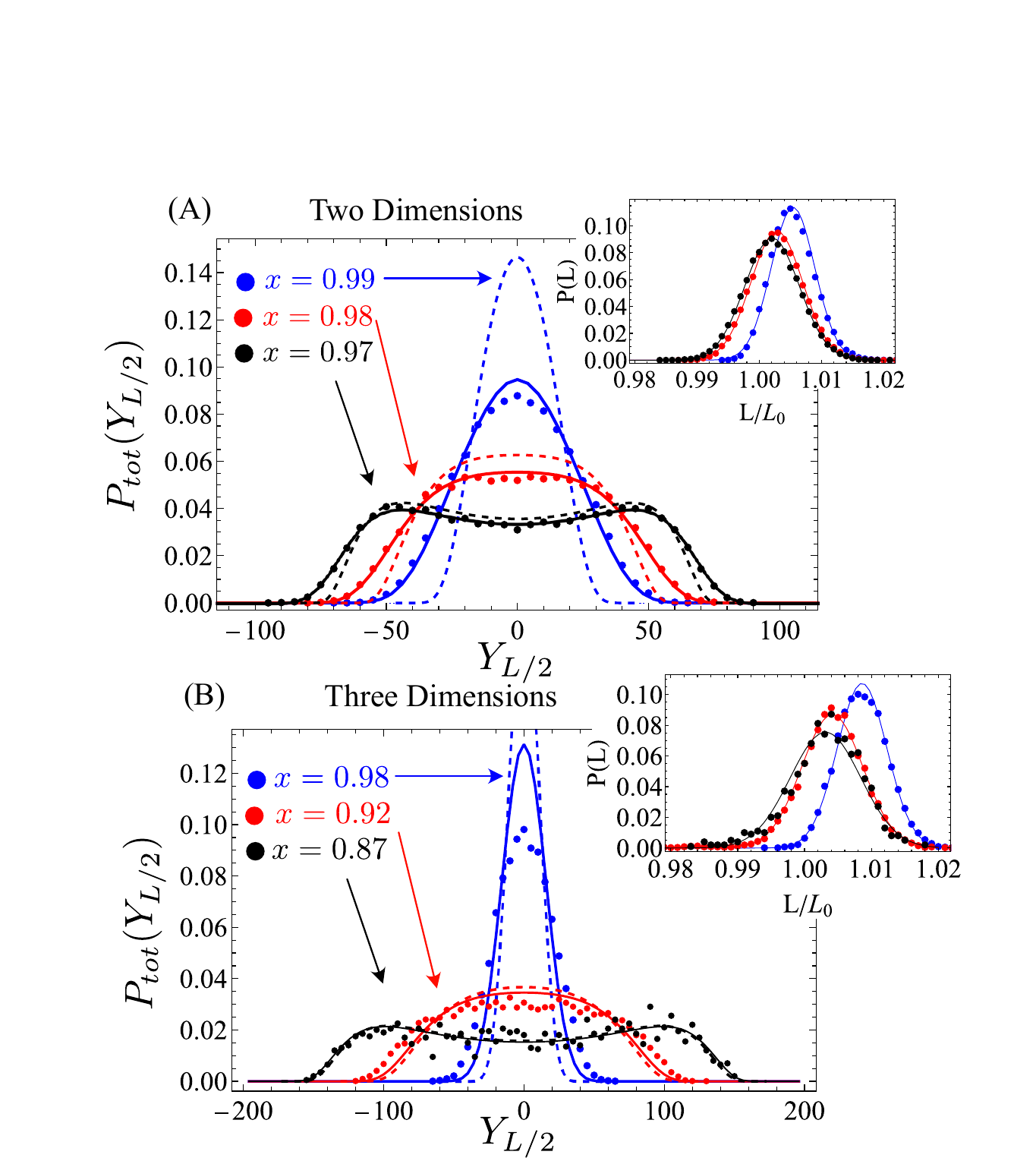}
\caption{\label{integrated.fig}  Simulated distributions with $\kappa_s/a^2=10k_BT$ (points) in comparison to the fixed length distributions in eq. \ref{distribution} (dashed lines) and eq. \ref{integratedDist} (solid lines) in two (A) and three (B) dimensions.  The insets show the distributions of the simulated lengths, which are well fit by a Gaussian.  The transition between the buckled and unbuckled states differs in two and three dimensions, and different values of compression are shown:  in (A) shown are $x=0.99$ (blue), 0.98 (red), and 0.97 (black), while in (B) shown are $x=0.98$ (blue), 0.92 (red), and 0.86 (black).  Note that in (B) the peak of the dashed blue line is cut off to improve visualization.}
\end{center}
\end{figure}

The distribution in eq. \ref{distribution} assumes a fixed $L$, but all biomolecules have a finite stretch modulus, and many computational models (including our MC algorithm) permit length fluctuations. For example, F-actin has a stretch modulus \cite{kojima1994direct} of $\approx 1.8 \times 10^9$ N/m$^2$ which allows for around $1-5 \%$ length fluctuations. In Fig. \ref{integrated.fig}, we show that the theoretical prediction in Eq. \ref{distribution} (dashed lines) in two and three dimensions fails to capture the simulated distribution $P(Y_{L/2})$ for large $x$ (stretching), but appears to perform well for small $x$ (compression).  The two dimensional simulation includes 50,000 simulated configurations for each distribution and for three dimensions there are approximately 8400 simulated configurations for each distribution. For both compression and stretching, we see the simulated length of the chain, $L_{k}=\sum_{i=1}^{N-1}|\rv_{i+1}^k-\rv_i^k|$, varies by $\approx 2\%$ (insets of Fig \ref{integrated.fig}) and that the average length is increased when stretched (because the locations of the peaks of the length distributions increase with $x$) but not when compressed.   The length distributions are all well-fit by a normal distribution, $p(L_{sim})=\exp[(L_{sim}-\bar L)^2/2\sigma_L^2]/\sqrt{2\pi \sigma^2_L}$ (the solid curves in the inset) with $\bar L$ and $\sigma_L^2$ the mean and variance of the simulated length.  An exact calculation of $\bar L$ and $\sigma_L$ is difficult analytically due to the constraint of fixed distance in the $x$-direction, and we simply calculate them from the data directly.  In computing $\bar L$ and $\sigma_L$ we exclude extensions that occur less than 0.1\% of the data, due to rare events of high compression of the bonds when $x$ is small.

Initially, we tested to see if we could fit the simulated distributions using Eq. \ref{distribution} with $L$ and $l_p$ as fitting parameters, but found the fitted distributions still did not agree well (data not shown).   To better match the simulations, we  compute the total distribution accounting for length fluctuations
\begin{eqnarray}
P_{tot}(y;x)=\int_{L_{min}}^{L_{max}} dL_{sim}\ p(L_{sim})\ \frac{P(y L_{sim};x L_0)}{{{\mathcal{N}}}(L_{sim},L_0)}\label{integratedDist},
\end{eqnarray} 
with the Gaussian prior $P(L)$ in the insets of Fig. \ref{integrated.fig}, $L_0=(N-1)a$ the length of the chain for $\kappa_s\to\infty$, and the normalization ${{{\mathcal{N}}}(L_{sim},L_0)}=\int_{Y_{min}}^{Y_{max}} dY P(y L_{sim},x L_0)$.  This integral must be evaluated numerically. The bound values of L, $L_{min}$ and $L_{max}$, are chosen such that $\int_{-\infty}^{L_{min}}dLp(L)=\int_{L_{max}}^{\infty}dLp(L)=10^{-5}$ (ignoring low-probability lengths), and $Y_{max}=-Y_{min}=(L_{sim}/2)\sqrt{1-x^2}$.  The numerical integrations uses a 25-point Gaussian quadrature, and the resulting distributions are shown as the solid lines in Fig. \ref{integrated.fig}.   The fixed length distribution functions in Fig. \ref{distribution} badly fail to capture the simulated distributions for $x=0.99$ (2 dimensions, dashed blue lines in Fig \ref{integrated.fig}(A)) or 0.98 (3 dimensions, dashed blue lines in Fig \ref{integrated.fig}(B)), while the convolved distributions $P_{tot}$ show improved agreement.  There is still some difference between the theory and simulations for large $x$, which may be due to the failure of a Gaussian theory to accurately describe a strongly-stretched inextensible chain \cite{MikesPaper}.  In two dimensions, the convolved distribution in eq. \ref{integratedDist} agrees with almost perfectly with the simulations while the fixed length theory fails to capture the tails well.  In three dimensions, the agreement for the fixed length theory is nearly quantitative below the buckling transition, suggesting that theory can be reliably used for high compression in three dimensions.  The agreement between the fixed-length theory in eq. \ref{distribution} and the convolved distribution in eq. \ref{integratedDist} suggests that our predictions of the location of the buckling transition will be accurate even for systems with finite stretching modulus.

\section{Discussion and Conclusions}

In this paper, we have studied the statistics of semiflexible filaments compressed due to pinning at the endpoints (in the Helmholtz ensemble) to better understand the buckling of stiff polymers when compressed.  We find that there is a continuous phase transition for a pinned polymer in both two and three dimensions, transitioning from configurations that thermally fluctuate around the compression axis for high extensions ($x=X_{ee}/L\gtrsim x_*$) to thermal fluctuations about bent configurations for strong compression ($x<x_*$).  This transition does not precisely coincide with the change in sign of the mean force (although is very close in two dimensions), nor does it coincide with the point at which the compressive force experiences a local minimum.  For rigid chains all three of these points nearly coincide, consistent with the observations in ref. \cite{colloidbuckling} (with $L/l_p\approx 117$).   We predict that even fairly flexible chains may also buckle (with $L/l_p$ as low as  $0.1$ or 0.2 in two or three dimensions respectively), but the buckling transition will occur for a larger value of $x$ than sharp changes in the force.

The stiffness of many biopolymers plays an essential role in the structure and dynamics of compressed filamentous networks and bundles, as demonstrated both theoretically\cite{wang2019buckling,broedersz2014modeling} and experimentally\cite{bustamante2003ten,bathe2008cytoskeletal,cooper2000structure,blanchoin2014actin,claessens2006actin,lieleg2008transient,lieleg2010structure}.  The persistence length of these molecules are central to characterizing their behavior when compressed.  In this work, we have derived a relation between the mean-field persistence length obtained from our theory, and while we found that in three dimensions the very simple relation $l_1=3l_p/2$ agrees well with simulated data, modeling a two-dimensional buckled filament requires a nonlinear relationship between $l_1$ and $l_p$.   The novel two-dimensional end-to-end distribution function predicted here may be relevant studying the statistics of surface-bound filaments or networks.   We also found that the distributions are quite sensitive to variations in length for stretched filaments, and the theory does a relatively poor job of capturing the distributions for the normalized extension $X_{ee}\approx (N-1)a$.  In three dimensions, we find the theory and simulations agree well for smaller $x$ (closer to the buckling transition), and the inextensible theory is likely sufficient for understanding the statistics of filaments with finite stretching moduli such as F-Actin.  In two dimensions, we found that the inextensible theory only qualitatively captures the simulated distributions and one must account for length fluctuations explicitly to quantitatively describe the distributions.  Despite this sensitivity in the distributions, the compression at which buckling occurs predicted by the inextensible theory ($x_*=\sqrt{1-c_dL/l_1})$  is expected to be accurate in both two and three dimensions.

In this paper, we have pinned the endpoints of the filament in one dimension in this paper, and induced buckling by decreasing the end-to-end distance along that axis.  This imposes no constraint on the direction of the bonds at the endpoints, but in many biologically or experimentally relevant conditions, one might expect constraints on the statistics of the endpoints\cite{blanchoin2014actin, brangwynne2006microtubules, chaudhuri2007reversible, freedman2017versatile}.  These may include rigid constraints of clamped (both endpoints normal to a surface) or cantilevered (one end normal to a surface and the other free) ensembles, but other softer constraints might be appropriate depending on the system.   It is certainly possible to include rigid constraints at the endpoints by computing averages as $\langle \cdots\rangle_{clamped}=\langle(\cdots)\delta(\uv_0-\hat x)\delta(\uv_L-\hat x)\rangle_1$, and we readily find that $\Omega_{clamped}=d/2l_p(1-x^2-y^2/\sigma(1-\sigma))$ to leading order in $L$, identical to the value of $\Omega$ found for free ends.  Unsurprisingly, the mean field solution for the extensive contribution to the free energy is independent of the endpoint constraints in the limit as $L\to\infty$, and suggests the buckling transition will occur at the same value of $x$ regardless of the endpoint conditions.  However, a weakness in the the mean-field approach when studying such endpoint constraints is how to handle terms in the free energy involving $\delta$, which suppress endpoint fluctuations.  In the absence of fluctuations at the endpoints we expect $\delta\to\infty$.  Rigidly constraining the endpoints and minimizing the resulting free energy yields an expression for $\delta$ that is an unwieldy function of $x$ and $y$.  We find that $\delta\propto L$ under clamped conditions (consistent with the expectation of $\delta\to\infty$ since $L$ is assumed large), and also find that $\delta$ diverges when $\Omega$ diverges (when $y=\sqrt{(1-x^2)/\sigma(1-\sigma)}$).  The extensive scaling of $\delta$ significantly complicates the analysis on the mean field level.  While an interesting direction for future work would compare the mean field predictions in the clamped case to simulations and develop an analytically tractable functional form, significant numerical work and simulations are required to better understand the effect of rigid bond boundary constraints.  

\section{Acknowledgements}
We are grateful to the members of the active-cytoskeleton focus group at the Center for Theoretical Biological Physics (CTBP) for stimulating discussions on a broad range of topics involving modeling of F-actin. AM also benefited greatly from feedback from her graduate committee members for this project. This work was completed in part with resources provided by the Research Computing Data Core at the University of Houston. We acknowledge funding from the National Science Foundation, NSF-PHYS-2019745 supporting this work, as well as computational resources through NSF-CNS-1338099.
\bibliography{buckling}
\bibliographystyle{unsrt}

\end{document}